\documentclass{article}
\usepackage{iclr2025_conference,times}

\usepackage[utf8]{inputenc} %
\usepackage[T1]{fontenc}    %
\usepackage{graphicx}
\usepackage{subcaption}
\usepackage{ulem}
\usepackage{amsmath}
\usepackage{mathtools}
\usepackage{hyperref}       %
\usepackage{wrapfig}
\setcitestyle{round}
\hypersetup{
   colorlinks=true,
   linkcolor=[RGB]{30, 30, 180},
   citecolor=[RGB]{30, 30, 180},
   urlcolor=magenta,
   pdfborder=0 0 0,
   pdftitle={},
   pdfsubject={}, 
   pdfkeywords={},
   pdfauthor={},%
   pdfstartview=FitH
}
\usepackage[capitalize,noabbrev]{cleveref}
\usepackage[nolist,nohyperlinks]{acronym}

\author{Gianluca Galletti$^{1}$,
  Fabian Paischer$^{1}$,
  Paul Setinek$^1$, 
  \textbf{William Hornsby}$^{2}$, 
  \textbf{Lorenzo Zanisi}$^{2}$,\\ 
  \textbf{Naomi Carey}$^{2}$ 
  \textbf{Stanislas Pamela}$^2$ 
  \textbf{Johannes Brandstetter}$^{1,3}$ \\
  {$^1$~ELLIS Unit, LIT AI Lab, Institute for Machine Learning, JKU Linz, Austria}\\
  {$^2$~UKAEA (United Kingdom Atomic Energy Authority), Culham Campus, Abingdon}\\
  {$^3$~Emmi AI GmbH, Linz, Austria}\\
  \texttt{paischer@ml.jku.at}
}

\usepackage{array}
\usepackage{amsmath}
\usepackage{amssymb}
\usepackage{mathtools}
\usepackage{amsthm}
\usepackage{bm}

\newcommand{\ourmethod}{5D Swin-UNet}
\newcommand{\ndwinmsa}{nDWin-MSA}
\newcommand{\ndswinmsa}{nDSWin-MSA}

\newcommand{\cG}{\mathcal{G}}

\newcommand{\cU}{\mathcal{U}} \newcommand{\cV}{\mathcal{V}}

\makeatletter
\newcommand{\dlmf}[1]{%
\citep[%
  \def\nextitem{\def\nextitem{, }}%
  \@for \el:=#1\do{\nextitem\href{http://dlmf.nist.gov/\el}{(\el)}}%
]{olver_nist_2010}%
}
\makeatother

\newcommand{\stoptocwriting}{%
  \addtocontents{toc}{\protect\setcounter{tocdepth}{-5}}}
\newcommand{\resumetocwriting}{%
  \addtocontents{toc}{\protect\setcounter{tocdepth}{\arabic{tocdepth}}}}

\makeatletter
    \newcommand{\settitle}{\@maketitle}
\makeatother

\begin{acronym}[LONGEST]
  \acro{ITG}{Ion Temperature Gradient}
\end{acronym}

\title{5D neural surrogates for nonlinear gyrokinetic simulations of plasma turbulence}
\iclrfinaltrue

\begin{document}
\pagestyle{fancy}

\stoptocwriting
\maketitle

\begin{abstract}
    Nuclear fusion plays a pivotal role in the quest for reliable and sustainable energy production. 
    A major roadblock to achieving commercially viable fusion power is understanding plasma turbulence, which can significantly degrade plasma confinement. 
    Modelling turbulence is crucial to design performing plasma scenarios for next-generation reactor-class devices and current experimental machines. 
    The nonlinear gyrokinetic equation underpinning turbulence modelling evolves a 5D distribution function over time. 
    Solving this equation numerically is extremely expensive, requiring up to weeks for a single run to converge, making it unfeasible for iterative optimisation and control studies. 
    In this work, we propose a method for training neural surrogates for 5D gyrokinetic simulations.
    Our method extends a hierarchical vision transformer to five dimensions and is trained on the 5D distribution function for the adiabatic electron approximation.
    We demonstrate that our model can accurately infer downstream physical quantities such as heat flux time trace and electrostatic potentials for single-step predictions two orders of magnitude faster than numerical codes.
    Our work paves the way towards neural surrogates for plasma turbulence simulations to accelerate deployment of commercial energy production via nuclear fusion.
\end{abstract}

\section{Introduction}
\label{sec:intro}

Turbulence is a key driver of plasma confinement degradation, as it causes Plasma to diffuse towards the reactor wall, resulting in most of the heat and particle transport in magnetic fusion devices, such as Tokamaks.
The growing turbulence is dampened by the zonal flow system, which regulates turbulence to reach a quasi-stationary saturated state \citep{Itoh_ZF_06}.
The design and control of performing plasma scenarios strictly requires knowledge of the turbulent transport in the saturated state.
This can be obtained via complex nonlinear gyrokinetic simulations, which involve evolution of a 5D distribution function over time.

Computationally affordable reduced-order quasilinear models of turbulent transport, such as QuaLiKiz \citep{Bourdelle2015,Citrin2017} and TGLF \citep{Staebler2007,Staebler2010}, are routinely adopted. 
They make simplifying assumptions about the mechanism of turbulence saturation and modality expressed as so-called saturation rules. %
As different saturation rules are fitted to specific turbulence modalities, a \textit{general} quasilinear model is not available to date, which limits their usefulness.
A fast, general and reliable estimate of turbulent fluxes is only attainable via expensive nonlinear gyrokinetic simulations.

Neural surrogates can reduce the complexity of nonlinear gyrokinetic simulations.
Current approaches operate on reduced input spaces \citep{narita_toward_2022,mitsuru_multimodal_2023}, which do not model the 5D distribution function. 
This results in information loss, as they do not capture interactions across all dimensions.
In this work, we attempt for the first time to train a neural surrogate for the evolution of the 5D distribution function.
To this end, we first collect simulation data using a state-of-the-art nonlinear solver, namely GKW \citep{PEETERS_GKW_2009}, run with adiabatic electrons for different values of \ac{ITG}.
To compress the high-dimensional input into a compact representation, we extend a hierarchical vision transformer to process the 5D input.
We call our method \ourmethod{} and train it to evolve the 5D distribution function over time.

Our neural surrogate accurately models physical quantities for single-step predictions.
To verify this, we integrate the predicted 5D distribution function to infer physical quantities such as heat flux time trace and electrostatic potentials for single-step predictions of \ourmethod{}.
We find that the predicted quantities align well with the ground truth for a simulation with a value for the ion temperature gradient (\ac{ITG}) which was never observed during training, except for an overestimation of the zonal flow.
Furthermore, \ourmethod{} is approximately two orders of magnitude faster than GKW for the adiabatic electron approximation.
In summary, our contributions are as follows.
\begin{itemize}
\item We provide a simple 2D visualization for the 5D distribution function.
\item We introduce \ourmethod{}, a hierarchical vision Transformer to compress, process and reconstruct 5D data, without convolutions.
\item We propose a physics evaluation to validate that our surrogate models accurately capture turbulent transport in Plasma.
\end{itemize}

\section{Related Work}

\textbf{Machine Learning for Gyrokinetics.}
Most of the literature to date has focused on multilayer perceptrons as surrogate models of turbulence models that adopt the quasilinear approximation. 
Faster integrated models \citep{aJINTRAC-Romanelli, ASTRA}  of tokamak discharges were obtained for interpretative modelling in existing machines \citep{Meneghini2017, plassche2020} as well as predictive modelling \citep{Citrin_2023} and control studies \citep{VanMulders2021} of future reactor-class devices, clearly highlighting the benefits of surrogate modelling, albeit on reduced order models.

The literature on surrogates for higher-fidelity models is sparse. 
To model the linear spectra of micro-tearing modes, \citet{Hornsby2024} propose leveraging gaussian process regression (GPR).
GPR also enables uncertainty quantification, but does not scale well to larger datasets and dimensionalities.
Therefore, \citet{narita_toward_2022} leverage convolutional neural networks based on the spectral images of the absolute values of the distribution function along with the electric potential at a fixed time slice and predict the corresponding heat flux and the time to saturation. 
\citet{mitsuru_multimodal_2023} extends this idea to a two-dimensional multimodal input space, including electrostatic potentials.
Our method fundamentally differs in that it directly models the time evolution of the 5D distribution function, thus enabling the computation of turbulent fluxes at any time. 

\textbf{Neural Operators.} 
Importantly, our \ourmethod{} does not fall into the category of neural operators, as it is not resolution invariant but operates on a fixed resolution.
However, in future work we aim to extend our model to a neural operator, hence we elaborate on important related work in this regard.
Neural operators \citep{lu_learning_2021,li_neural_2020, kovachki_neural_2023,alkin_upt_2024} are formulated with the objective of learning a mapping between function spaces, usually defined as Banach spaces U, V of functions defined on compact input and output domains X and Y, respectively. 
Neural operators enable continuous outputs that remain consistent across varying input sampling resolutions. 
A neural operator $\hat{\cG} : \cU \mapsto \cV$ approximates the ground truth operator $\cG : \text{U} \mapsto \text{V}$, and is usually composed of three maps $\hat{\cG} := \text{D} \circ \text{A} \circ \text{E}$ \citep{seidman_nomad_2022}, comprising the encoder E, the approximator A, and the decoder D. 
Training a neural operator involves constructing a dataset of input-output function pairs evaluated at discrete spatial locations. 
Training minimizes a mean squared error loss over the discretized space using gradient descent.

Over the recent years, driven by advances in neural operator learning, deep neural network-based surrogates have emerged as a computationally efficient alternative in science and engineering \citep{thuerey_physics_2021,zhang_artificial_2023,brunton_machine_2020}, impacting e.g., weather forecasting \citep{thorsten_fourcastnet_2023,bi_accurate_2023,lam_learning_2023,nguyen_climax_2023,bodnar_aurora_2024}, protein folding \citep{jumper_highly_2021,abramson_accurate_2024}, material
design \citep{merchant_scaling_2023,zeni_generative_2025,yang_mattersim_2024}, and multi-physics modeling~\cite{alkin2024neuraldem}. All these success stories share the common thread of deep learning surrogates unlocking new possibilities to overcome seemingly insurmountable challenges~\citep{brandstetter2024envisioning}.

\textbf{Swin Transformers.}
Other works, such as the 4D fMRI Swin Transformer \citep[SwiFT]{kim2023swift} extend Swin Transformers to the specific case of four dimensions (three spatial and time) to process fMRI scans. 
Since our nD layers do not rely on assumptions about the number of dimensions, we effectively automate the process used to derive SwiFT (4D) and Video-Swin \citep{liu_video_2022} (3D) from standard 2D Swin layers.

\section{Methods}
\begin{figure}[t]
    \centering
    \hspace*{1em}
    \begin{minipage}{0.5\linewidth}
        \centering
        \includegraphics[height=170px]{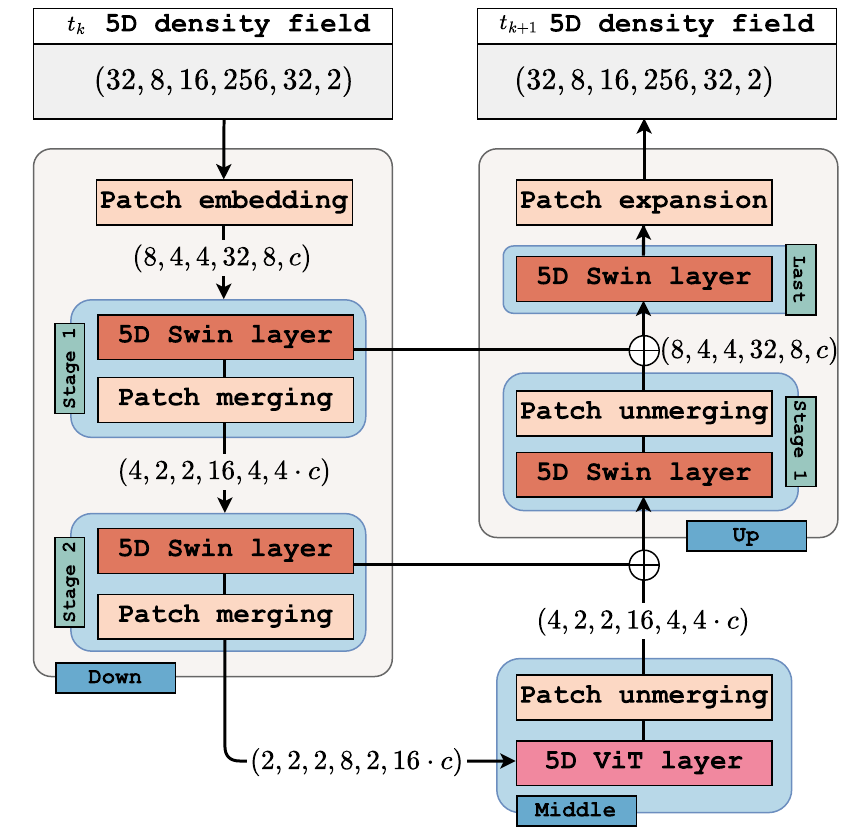}
    \end{minipage}%
    \begin{minipage}{0.5\linewidth}
        \centering
        \includegraphics[height=170px]{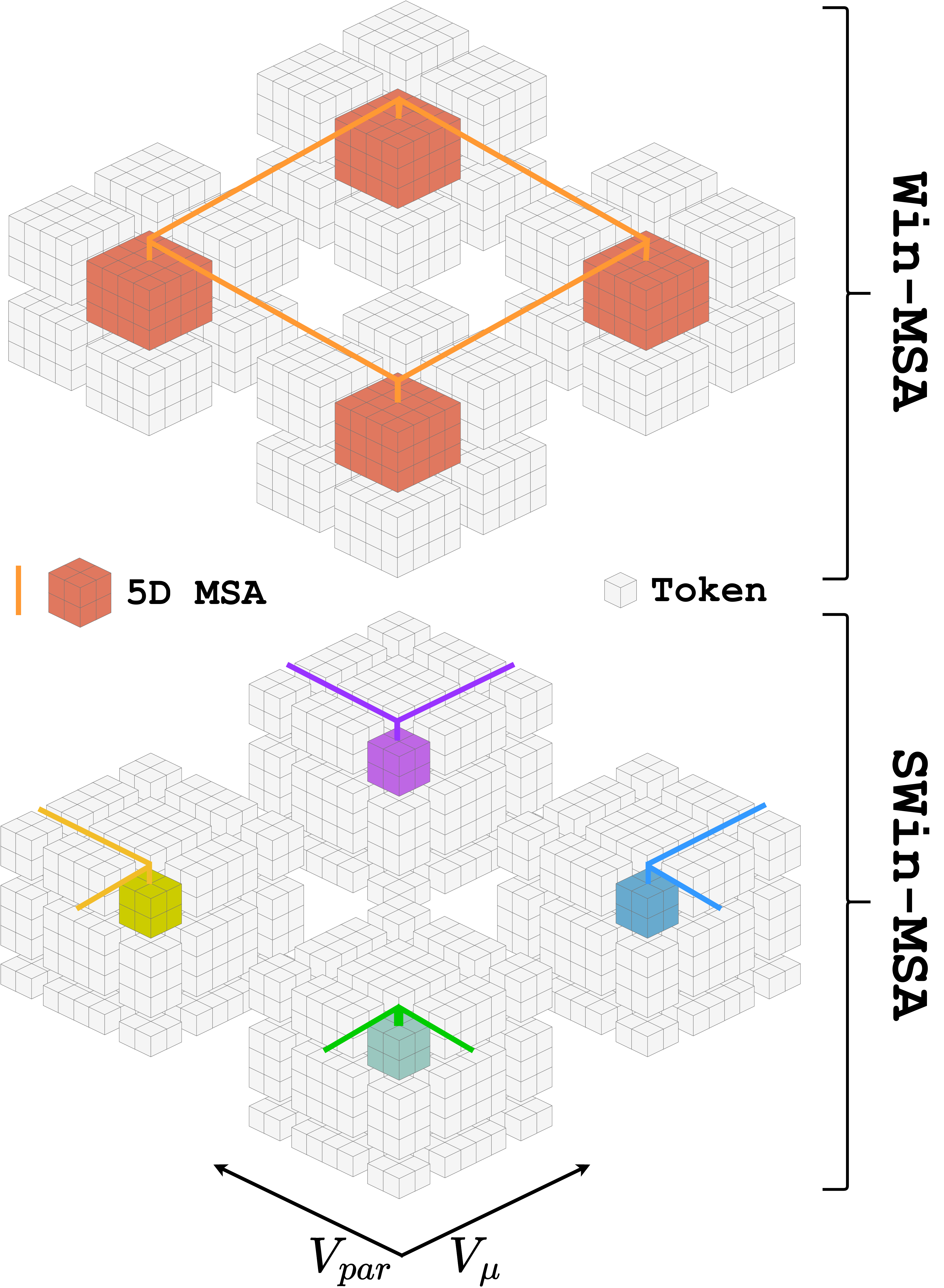}
    \end{minipage}
    \caption{Overview of the model architecture and n-dimensional attention. \textbf{Left:} Our \ourmethod{} with two down/upsampling stages. 
    We indicate shapes at each resolution change. \textbf{Right:} The locality of the n-dimensional shifted window attention for \ndwinmsa{} (top) and \ndswinmsa{} (bottom), illustrated as a 2D plane of 3D window-partitioned tokens. Connected components are highlighted as colored tokens in the 3D blocks, and as lines connecting them across dimensions.}
    \label{fig:overview}
\end{figure}

First, we elaborate on the data generation, preprocessing, and visualization techniques in \cref{subsec:data}.
Next, we explain our proposed \ourmethod{} and elaborate on the architectural details in \cref{subsec:architecture}.
Finally, in \cref{subsec:evaluation}, we elaborate on our evaluation protocol.

\subsection{Data Generation}
\label{subsec:data}

Gyrokinetic simulations are usually initialised from noise and evolved according to nonlinear gyrokinetic equations (\cref{gyrovlas}). 
The simulations usually follow a certain pattern.
In the linear phase the linear modes start growing, resulting in an initial increase in the heat flux time trace.
Afterwards, the simulations enter the nonlinear or saturated regime where the linear modes start interacting, resulting in an oscillatory behaviour of heat flux.
Turbulence is generally established in the nonlinear regime and complex integrals over long time intervals are required to obtain an estimate for the heat flux.

We rely on a nonlinear direct numerical gyrokinetic solver to produce samples of the 5D distribution function, namely GKW \citep{PEETERS_GKW_2009}.
For background on the gyrokinetic framework, we refer the reader to \cref{app:gyrokinetics}.
We consider the adiabatic electron case, where the velocity distribution of electrons is assumed to be Boltzmann-like. 
Therefore, only the distribution function of the ions is considered.
To obtain different trajectories, we vary the \ac{ITG} and collect 5D fields of resolution $(32 \times  8 \times 16 \times 255 \times 32 \times 2)$ where the sixth dimension corresponds to the real and imaginary part of the ballooning transform, which is usually used to describe plasma coordinates. 
The five remaining spatial dimensions are denoted as $(V_{||} \times V_{\mu} \times s \times k_x \times k_y)$, where $V_{||}$ and $V_{\mu}$ represent velocities parallel and perpendicular to the field lines, $s$ is the toroidal angle and $k_x$ and $k_y$ are spatial coordinates in the spectral space.
We run GKW for different values of \ac{ITG} and dump the distribution function every 20 steps to ensure sufficient change.
The collected trajectories have 500 such steps that are additionally subsampled every third, resulting in an overall time-coarsening of 60 times and a total of 166 snapshots per trajectory.
The dataset comprises five trajectories, resulting in 660 training samples and one unseen trajectory (165 samples).
\cref{app:data_generation_visualisation} provides a detailed discussion on the choice of \acp{ITG} used for data generation.

\begin{figure}
    \centering
    \begin{minipage}{.45\textwidth}
        \centering
        \includegraphics[width=\textwidth]{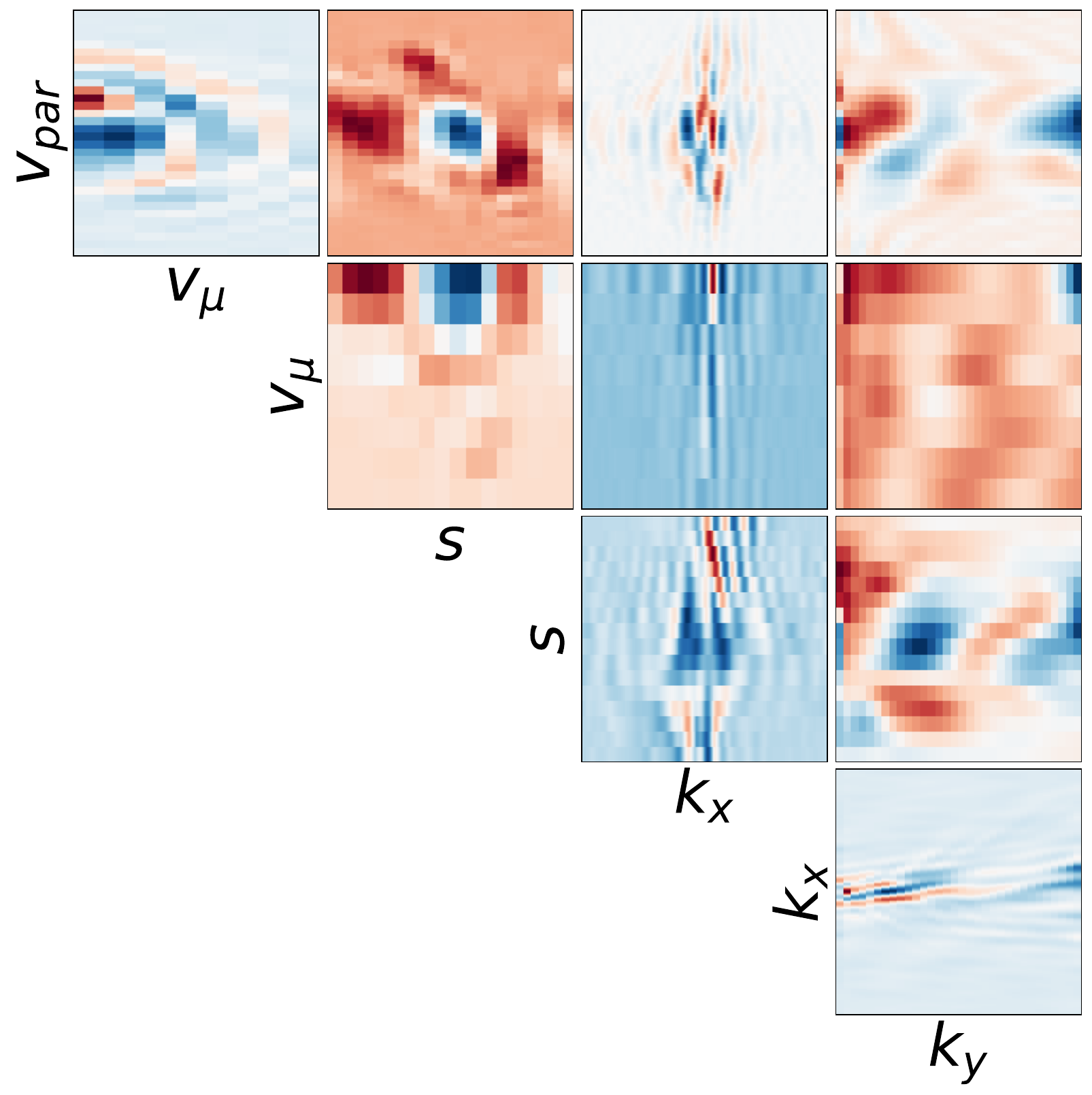}
    \end{minipage}
    \begin{minipage}{.45\textwidth}
        \centering
         \includegraphics[width=\textwidth]{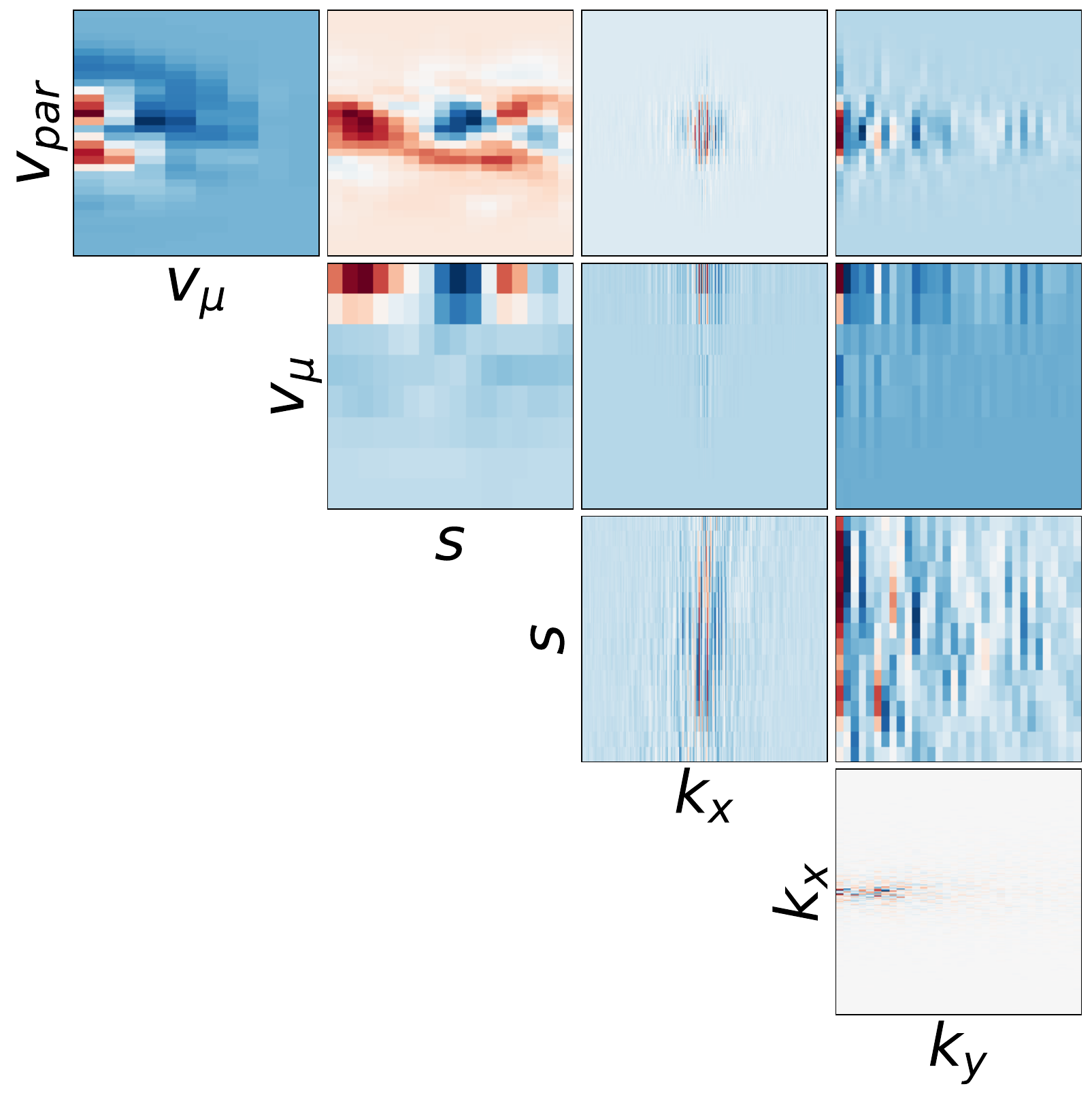}
    \end{minipage}
    \caption{\textbf{Left:} 5D distribution function with an \ac{ITG} of $7.9$ in the linear phase ($t=1.9$). \textbf{Right:} 5D distribution function with an \ac{ITG} of $7.9$ in the saturated phase ($t=50.6$).}
    \label{fig:deltaf_main}
\end{figure}

To visually inspect the data, we slice trajectories according to timesteps and visualize the 5D data as $\binom{5}{2}=10$ images of non-repeating combinations of the five different axes. 
For each combination of two axes, we either reduce or slice the remaining three axes to project the field down to an image. 
An example visualization for reduction via averaging for time slices of the linear and nonlinear phase of a simulation can be observed in \cref{fig:deltaf_main}.
The visualizations for the sliced samples are presented in \cref{fig:4x4_gt_t0} and \cref{fig:4x4_gt_t1} in \cref{app:data_generation_visualisation}.

\subsection{\ourmethod}
\label{subsec:architecture}

To process the 5D input, we generalize the hierarchical Swin Transformer \citep{liu_swin_2021} to an arbitrary number of dimensions.
The input to \ourmethod{} is a 5D field of shape $V_{par} \times V_{\mu} \times s \times x \times y \times 2$, which consists of the real and imaginary parts of a time snapshot of the nonlinear gyrokinetics simulation. 
Similarly to other conventional 2D Vision Transformers \citep{dosovitskiy_image_2021,liu_swin_2021}, we first partition the input into non-overlapping patches with an $n$-dimensional patch embedding layer, mapping patch-local information into tokens. 
Patches are then processed with a UNet-style architecture \citep{ronneberger_unet_2015}, using patch merging and unmerging to produce multiscale hierarchical representations. 
The original space is reconstructed from patch tokens using a patch expansion layer, which nonlinearly expands the patch space to the original resolution.
The proposed \ourmethod{} architecture is shown in Figure \ref{fig:overview}, left. 

The multi-head self-attention (MSA) used in Vision Transformers scales quadratically with the sequence length in the 2D case.
This effect is exacerbated in 5D, making MSA prohibitively expensive.
Therefore, we apply MSA only on local windows as in \citep[SWin]{liu_swin_2021}.
This is agnostic to dimensionality, i.e., it can be expanded to any number of dimensions with varying window sizes per dimension (\ndwinmsa).
To enable interaction between neighboring windows, we also extend cyclic shifts of SWin to 5D (\ndswinmsa).
Finally, we apply self-attention within each window in parallel.
Figure \ref{fig:overview} (right) visualizes the attention scope for one window in 5D for \ndwinmsa{} (top) and \ndswinmsa{} (bottom).

\subsection{Evaluation}
\label{subsec:evaluation}

To evaluate our method, we consider three approaches.
As a first evaluation, we visualize the outputs of the surrogate model.
This is done in the same manner as elaborated in \cref{subsec:data} and we apply this scheme to the ground truth data and to the model prediction to obtain a side-by-side comparison.
The second evaluation corresponds to visual inspection of the electrostatic potentials.
Potentials are computed by an integral over the velocities $V_{||}$ and $V_\mu$ of the 5D distribution function. 
on the predictions of our surrogate model.
Since electrostatic potentials are 3D fields, we slice them along the middle of the toroidal angle $s$ and visualize them as images.
Finally, we evaluate whether our surrogate model accurately predicts the heat flux time trace.
The heat flux is a scalar value that oscillates along an average in the saturated phase, and it is computed as a function of the 5D distribution function and the electrostatic potential.
It is a key quantity that is used in downstream integrated modelling tools \citep{aJINTRAC-Romanelli}.
This evaluation provides insight into whether the model accurately captures the underlying physics.

\section{Experiments}
\label{sec:experiments}

\begin{figure}[t]
    \centering
    \includegraphics[width=\textwidth]{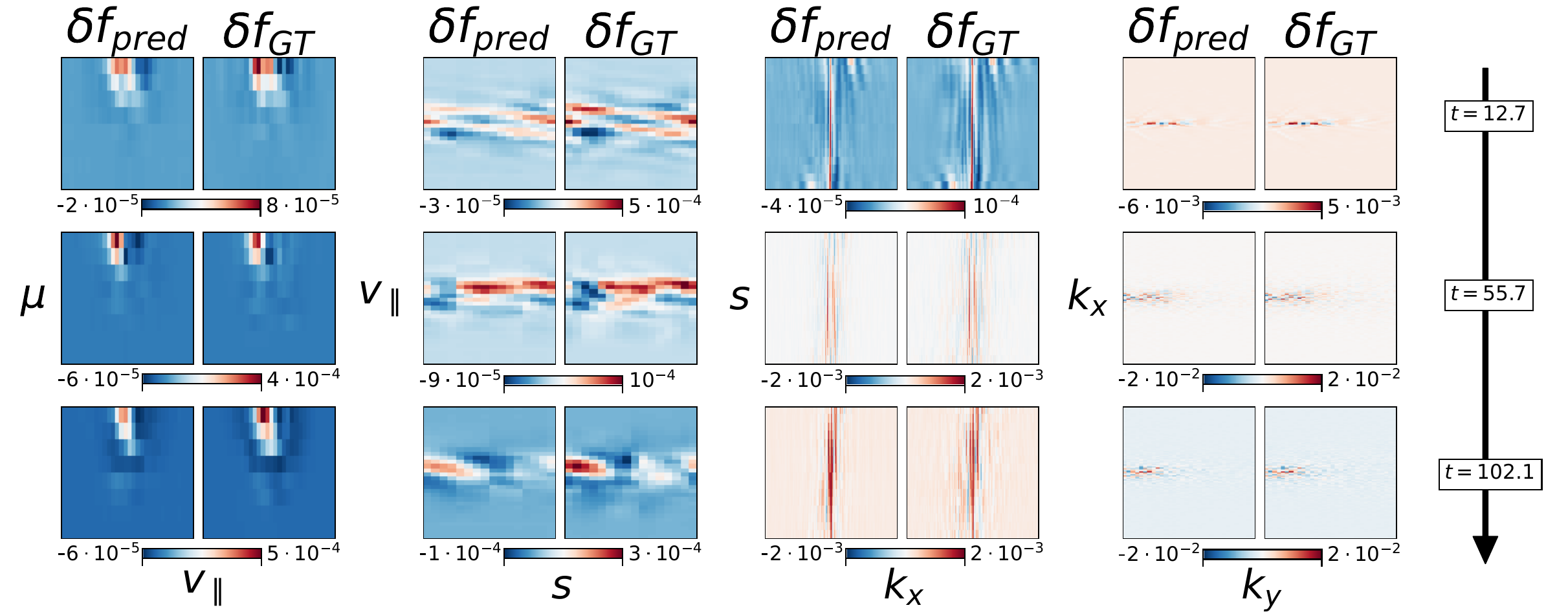}
    \caption{\ourmethod{} can accurately predict the distribution functon in five dimensions. We report one-step model predictions of the distribution function, $\delta f_{\text{pred}}$, versus ground truth, $\delta f_{\text{GT}}$, over time. For each axis pair, the corresponding 2D projections are obtained by averaging over the remaining dimensions.} %
    \label{fig:pred_df_poten}
\end{figure}

In our experiments, we train \ourmethod{} for next-step prediction of the nonlinear gyrokinetic simulation.
We split the data according to trajectories into training and validation.
Specifically, we evaluate on one holdout trajectory and use the remaining ones for training.
This way, we evaluate generalization to a holdout trajectory produced with a different \ac{ITG}.
Each sample of a trajectory is normalized on a per-sample basis.
We train our model for 100 epochs and evaluate every 20 epochs on the holdout trajectory.
For implementation details, we refer the reader to \cref{app:impl_details}.

\textbf{Distribution function.}
We visualize the predicted 5D field of the distribution function over time.
\cref{fig:pred_df_poten} shows averages for four of the 10 possible combinations of the different axes for three different timesteps.
The first timestep is from the linear regime, whereas the latter are from the saturated phase.
The model prediction generally aligns very well with the ground truth, as both model prediction and ground truth are displayed with euqal colour scale.

\begin{figure}[h]
    \centering
    \begin{subfigure}{0.48\textwidth}
        \centering
        \includegraphics[width=\textwidth]{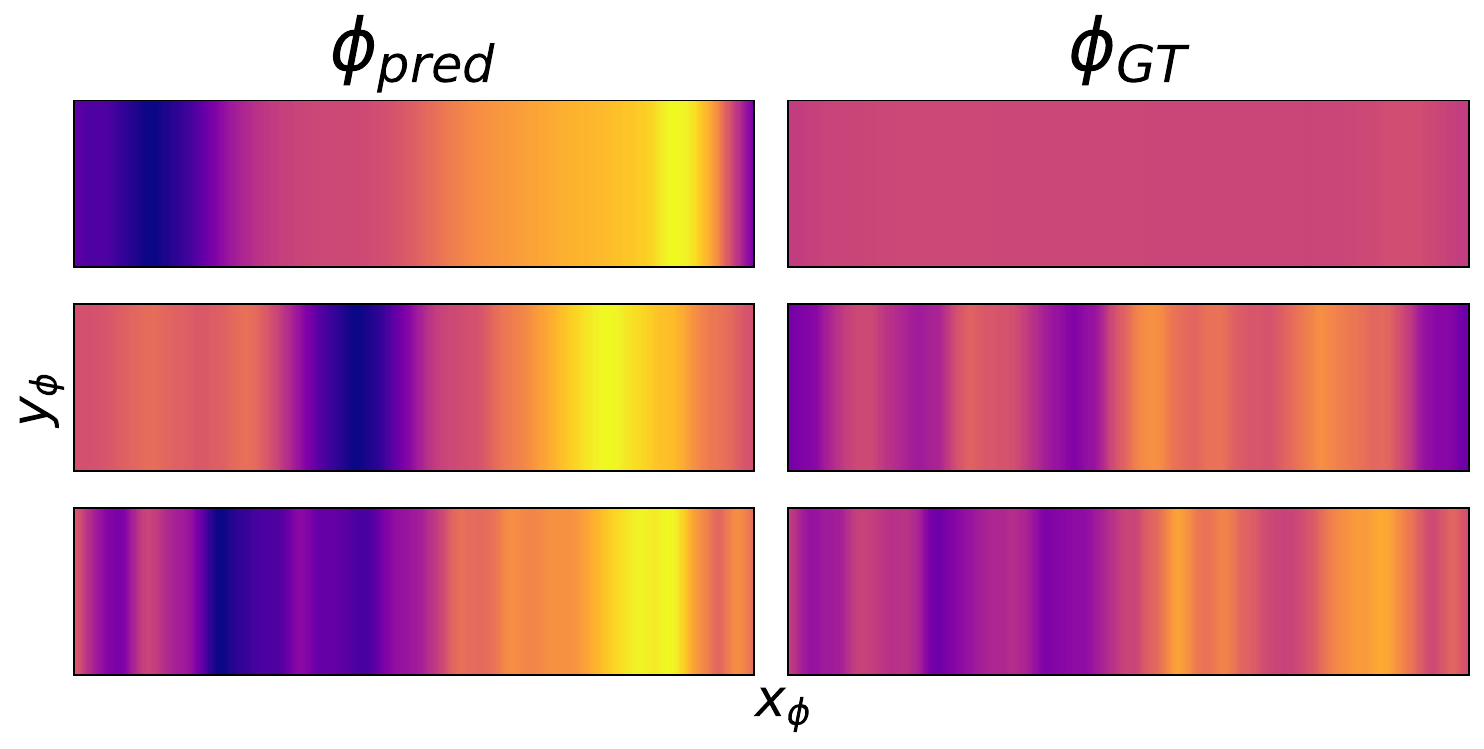}
        \caption{Zonal mode of the electrostatic potential.}
        \label{fig:pot_only_zonal}
    \end{subfigure}
    \hspace{0.02\textwidth}
    \begin{subfigure}{0.48\textwidth}
        \centering
        \includegraphics[width=\textwidth]{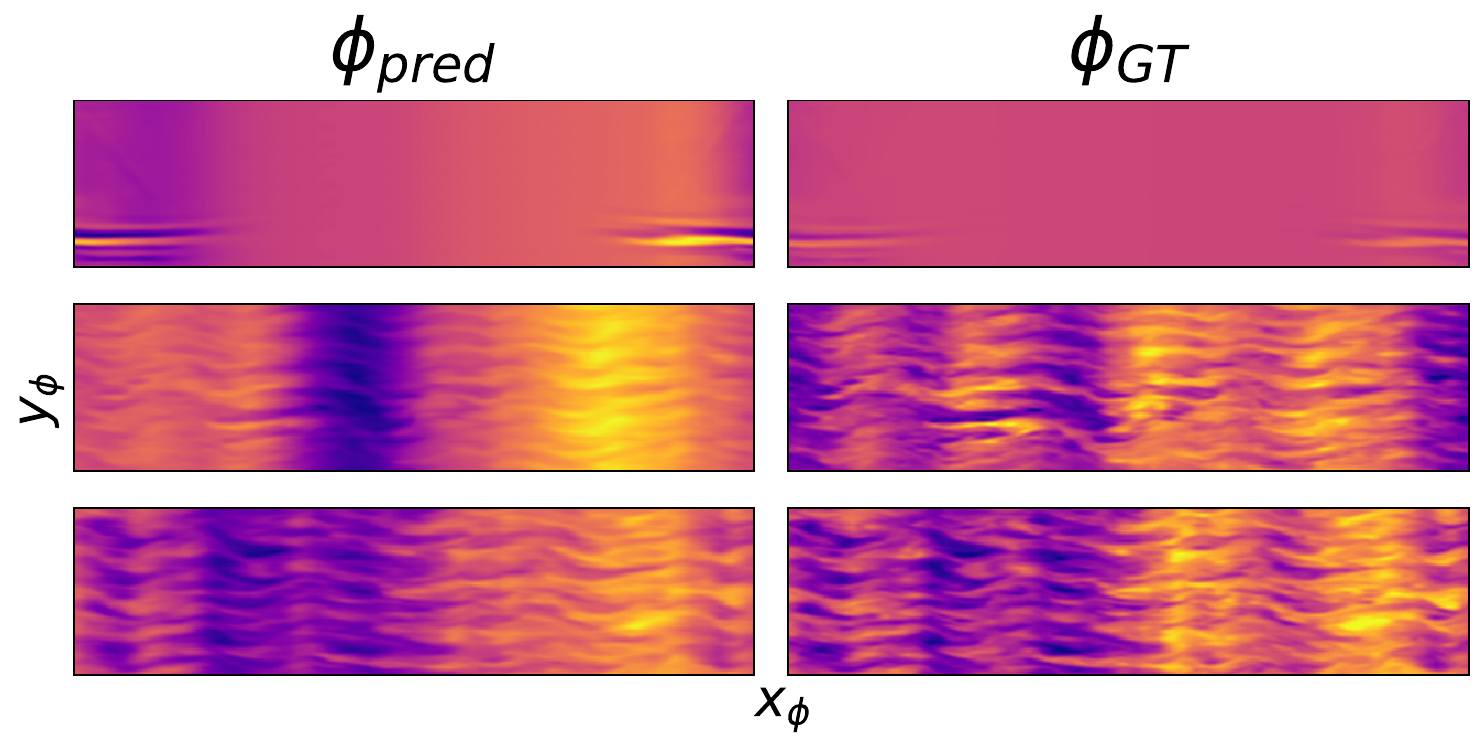}
        \caption{Electrostatic potential with zonal mode.}
        \label{fig:pot_zonal}
    \end{subfigure}
    \caption{Visualization and comparison of the electrostatic potentials for model predictions ($\phi_{\text{pred}}$) versus ground truth ($\phi_{GT}$). In accordance with \cref{fig:pred_df_poten}, rows correspond to timesteps $[12.7, 55.7, 102.1]$. \Cref{fig:pot_only_zonal} shows only the zonal component, while \cref{fig:pot_zonal} displays all modes of the field. While the vertical waves are well captured, the model overestimates the zonal flow, particularly at the first timestep, corresponding to the linear phase.}
    \label{fig:pot_app}
\end{figure}

\textbf{Electrostatic Potentials.}
We obtain the electrostatic potentials as described in \cref{subsec:evaluation} and provide a visual comparison of the predicted and ground-truth electrostatic potential in \cref{fig:pot_zonal}.
It shows that the the vertical wave vector is well reproduced, indicating that our surrogate model captures key aspects of the underlying physics.
However, there is a misalignment with respect to the first mode (zonal mode) that represents the zonal flow.
To further investigate this, we show the first mode of the electrostatic potential in \cref{fig:pot_only_zonal}.
Interestingly, \ourmethod{} strongly overestimates the presence of zonal flow in the linear phase where it is usually not present (top row).
This is likely caused by an imbalanced dataset, where the saturated phase -- with a stronger zonal mode -- is overrepresented.
The zonal flow does not directly contribute to the resulting heat flux in the saturated phase; however, it dampens the development of turbulent transport.
This is problematic, especially for autoregressive rollouts, where the predicted zonal flow in the linear phase results in a decaying heat flux.
Therefore, we aim to explore different sampling strategies or tailored losses emphasizing the zonal mode in the future.

\begin{wrapfigure}{r}{.5\textwidth}
    \vspace{-1em}
    \centering
    \includegraphics[width=\linewidth]{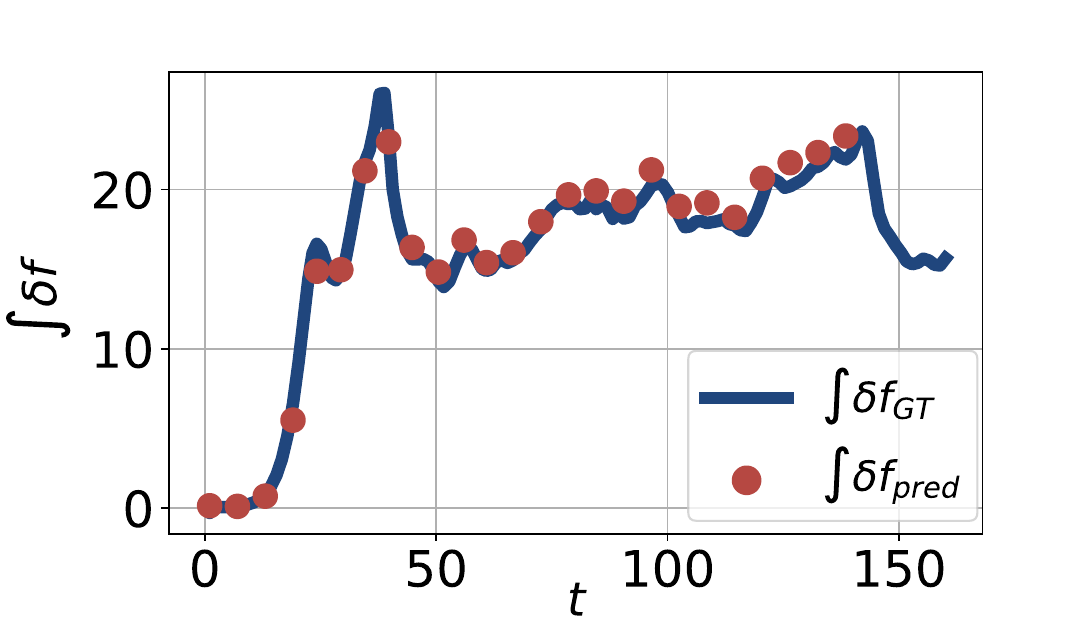}
    \caption{Heat flux time trace ($\int \delta f$) for ground-truth (GT) and single-step prediction of \ourmethod{} (pred) for the holdout trajectory (\ac{ITG}=6.9).}
    \label{fig:flux_time_trace}
\end{wrapfigure}

\textbf{Heat flux time trace.}
We assess whether the model correctly infers the heat flux time trace of the holdout simulation (\ac{ITG}=6.9).
\cref{fig:flux_time_trace} shows the predicted heat flux versus the ground truth.
The model predictions align very well with the ground-truth heat flux time trace.
This can be traced back to the vertical wave vector in the electrostatic potential (see \cref{fig:pred_df_poten}, right), which is successfully recovered by the surrogate model.
This indicates that \ourmethod{} successfully learns to capture parts of the underlying physics.
However, the zonal mode predicted by the surrogate is incorrect.
As zonal flow does not contribute to heat flux, it is not visible in \cref{fig:flux_time_trace}. 

\textbf{Efficiency.} One-step prediction only requires 360 ms on average for \ourmethod{} on a Nvidia A100 GPU, while the numerical solver GKW requires approximately 80s using 64 cpu cores.
Therefore, \ourmethod{} is around two orders of magnitude faster than GKW.

\section{Conclusions and future work}

We present a neural surrogate model for nonlinear gyrokinetic equations modelling turbulent transport in Plasmas.
We first present a simple recipe to visually inspect the 5D distributon function.
Then, we propose a neural surrogate that operates in 5D space, namely \ourmethod, and demonstrate that it accurately captures the underlying physics for single-step prediction.
Our work paves the way towards neural surrogates for 5D turbulence modelling for nuclear fusion.

As a first step for future work we plan to incorporate additional diagnostics to verify that the surrogate model accurately captures physics beyond electrostatic potentials.
Furthermore, a fruitful direction is to include a separate head for explicit prediction of electrostaic potentials. This alleviates the requirement for expensive GKW calls, needed for computing the potential field and the heat flux integrals.
Our aim is to achieve stable long-term autoregressive rollouts with our surrogate model which is currently hindered by the overemphasis on the zonal flow.
Eventually, our vision is to extend this work beyond next-step prediction to neural operators \citep{lu_learning_2021,li_neural_2020,kovachki_neural_2023,alkin_upt_2024} and enable transfer from lower-fidelity simulations that can be produced more efficiently to higher-fidelity ones that can take up to weeks to produce and are of different resolution.

\bibliographystyle{iclr2025_conference}
\bibliography{references}

\appendix
\section*{Supplementary Material}
\resumetocwriting

\title{Supplementary Material}
\settitle

\renewcommand{\baselinestretch}{0.1}\normalsize
\tableofcontents
\renewcommand{\baselinestretch}{1.0}\normalsize

\section{Gyrokinetic framework}
\label{app:gyrokinetics}

Solved within GKW is the gyrokinetic set of equations.  
The full details can be found in \cite{PEETERS_GKW_2009} and references therein. 
The $\delta f$ approximation is used, in which the distribution function is split into a background, Maxwellian distribution function $F_{M}$, and a perturbed distribution $f$ which is a 5 dimensional function, $f = f(k_{x},k_{y},s, V_{||}, V_\mu)$.
The final equation for the perturbed distribution function $f$, for each species can be written in the form 
\begin{equation}
\frac{\partial f}{\partial t} + (v_\parallel {\bf b} + {\bf v}_D) \cdot \nabla f +  {\bf v}_\chi\cdot \nabla f  
-\frac{\mu B}{m} \frac{\textbf{B} \cdot \nabla B}{B^2} \frac{\partial f}{\partial v_\parallel} = S, 
\label{gyrovlas}
\end{equation}
where $S$ is the source term which is determined by the background distribution function, $\mu$ is the magnetic moment, $v_{||}$ is the velocity along the magnetic field, ${\bf v}_D$ denotes the sum of the drift velocities, ${\bf v}_\chi$ is the $E\times B$ velocity, $B$ is the magnetic field strength, m and Z are the particle mass and charge number respectively.  The background is assumed to be a shifted Maxwellian ($F_M$), whose terms are moved to the right-hand side as a source term $S$, which represent the drives of the underlying instabilities that are studied. The electrostatic potential, $\phi$, is calculated from the gyrokinetic Poisson equation.

The thermal velocity $v_{\rm th}\equiv \sqrt{ 2 T / m}$, and the major radius ($R$) are used to normalise the length and time scales.
Using standard gyrokinetic ordering, the length scale of perturbations along the field line ($R \nabla_\parallel  \approx 1$) is significantly longer than those perpendicular to the field ($R \nabla_\perp \approx 1/ \rho_*$). Here, $\rho_* = \rho_i / R$ is the normalised ion Larmor radius (where $\rho_i = m_i v_{th} / e B$ and $v_{th} = \sqrt{2 T_i / m_i}$).  To harness this, GKW uses straight field-line Hamada coordinates ($s,\zeta,\psi$) where $s$ is the coordinate along the magnetic field and $\zeta$ is the generalised toroidal angle. 
GKW uses a Fourier representation in the toroidal ($y$) and radial directions ($x$), perpendicular to the magnetic field.

\section{Data Generation and Visualisation}
\label{app:data_generation_visualisation}

\textbf{Data Generation.} 

\begin{figure}[t]
    \centering
    \includegraphics[width=0.8\textwidth]{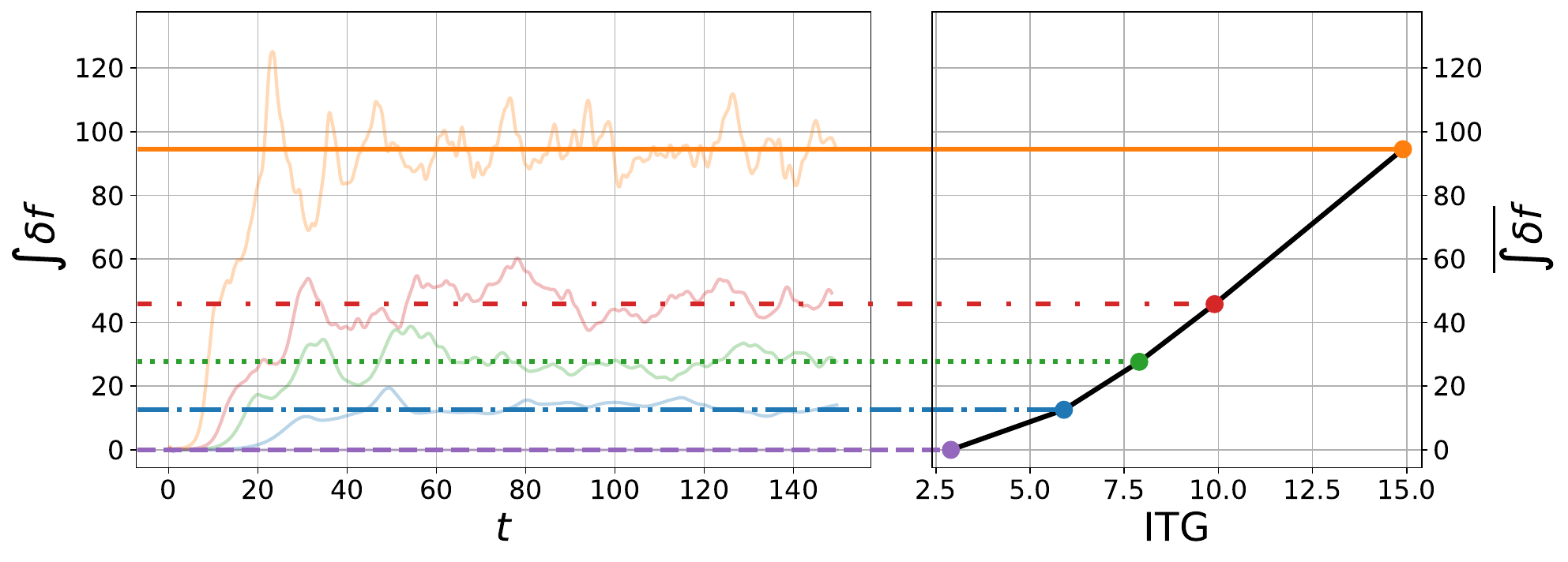}
    \caption{Flux trace over time (left) and averaged (right) for selected \ac{ITG} values.}
    \label{fig:itg}
\end{figure}

The choice of \ac{ITG} significantly impacts the behaviour of the simulations as it is a major driver of plasma turbulence.
In general, higher values lead to the accumulation of particle density, resulting in drift effects that eventually turn into turbulence. 
This can be visualized by comparing flux traces across different \acp{ITG} (see \cref{fig:itg}).
Generally, as \ac{ITG} increases, the system exhibits more nonlinear interactions, leading to higher heat flux values, $\int \delta f$.
For values of $\{5.9, 7.9, 10, 15\}$, one can clearly separate between a linear and a saturated phase; this distinction is not evident for lower values.
As values above $15$ are usually not observed in Tokamaks and those below $3$ do not cause any turbulent behaviour at all (see violet curve in \cref{fig:itg}), we sample within the range $[4, 10]$ to collect the first dataset.
Specifically, we use $\text{ITG} \in [4.9, 5.9, 6.9, 7.9, 8.9]$, resulting in a training set of 4 trajectories with 660 samples and a validation set of 165 samples.
Stored with single precision, this leads to a dataset size of 456 GB.

\begin{figure}[ht!]
    \centering
    \begin{subfigure}{0.45\textwidth}
        \includegraphics[width=\textwidth]{figs/5Dplots/compact_avg_7.9_1.9.pdf}
        \caption{Mean}
        \label{fig:gt_mean_t0}
    \end{subfigure}
    \hspace{0.05\textwidth}
    \begin{subfigure}{0.45\textwidth}
        \includegraphics[width=\textwidth]{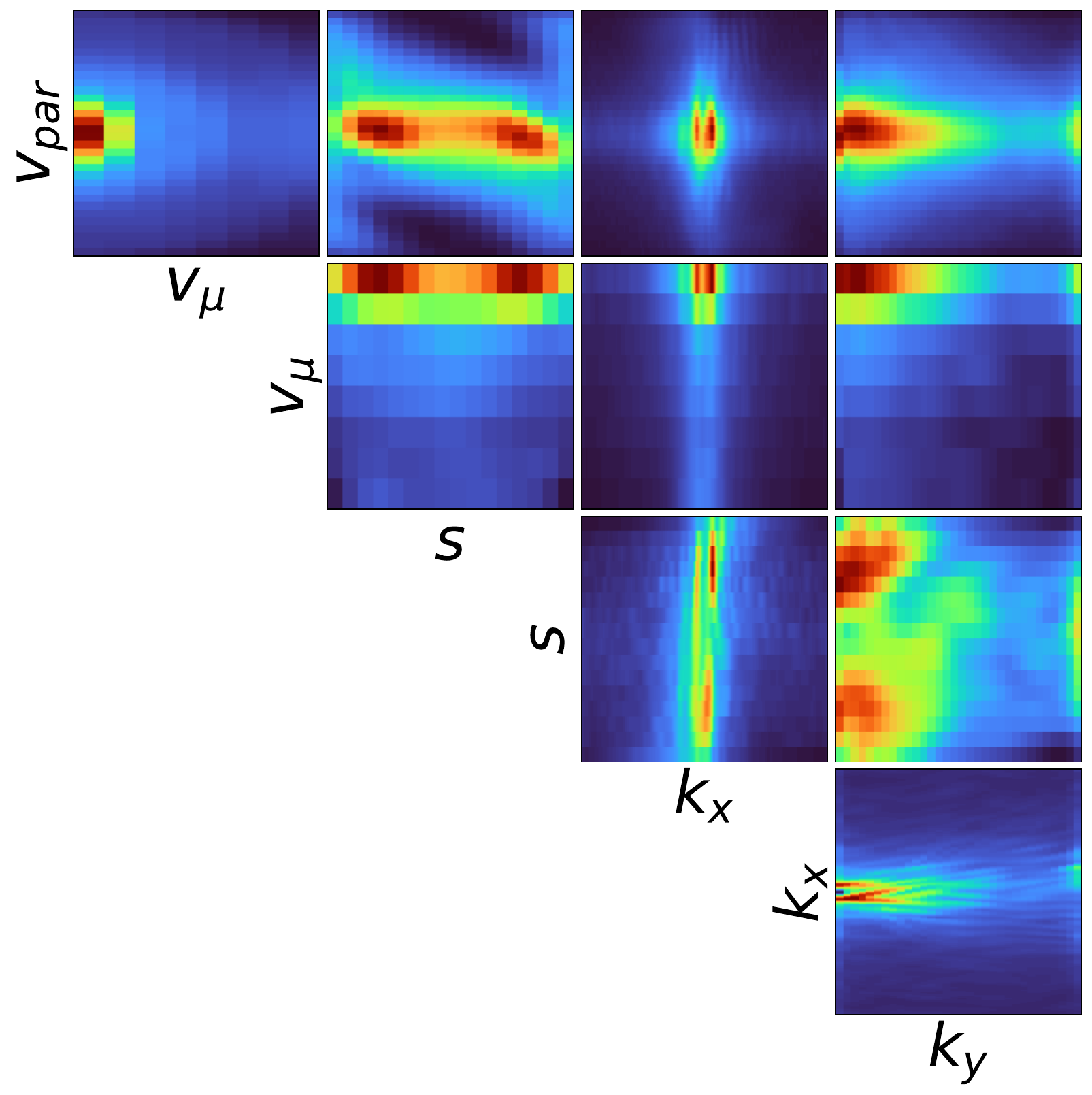}
        \caption{Standard Deviation}
        \label{fig:gt_std_t0}
    \end{subfigure}
    \par\bigskip %
    \begin{subfigure}{0.3\textwidth}
        \includegraphics[width=\textwidth]{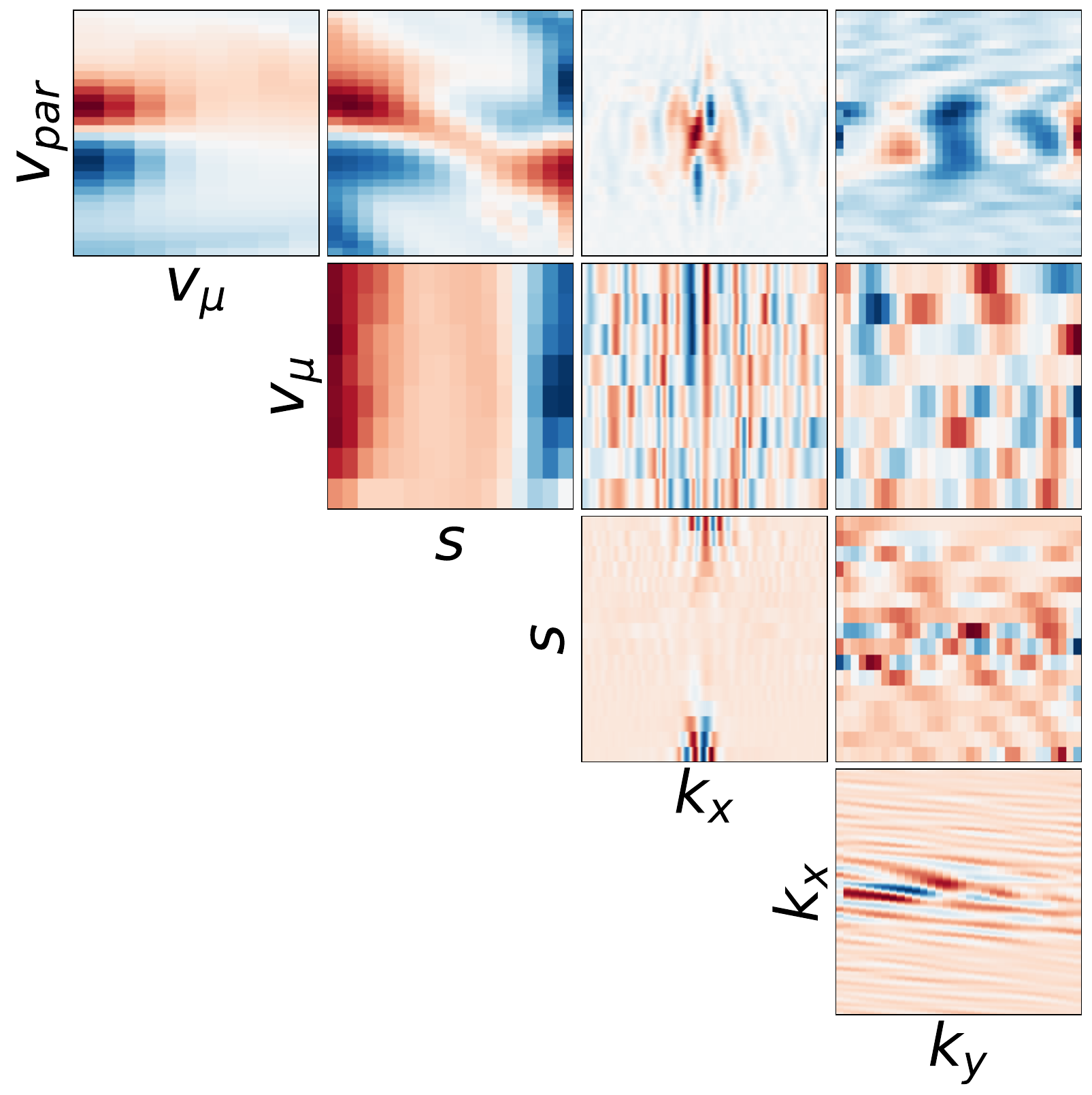}
        \caption{Start slice}
        \label{fig:gt_start_t0}
    \end{subfigure}
    \hspace{0.01\textwidth}
    \begin{subfigure}{0.3\textwidth}
        \includegraphics[width=\textwidth]{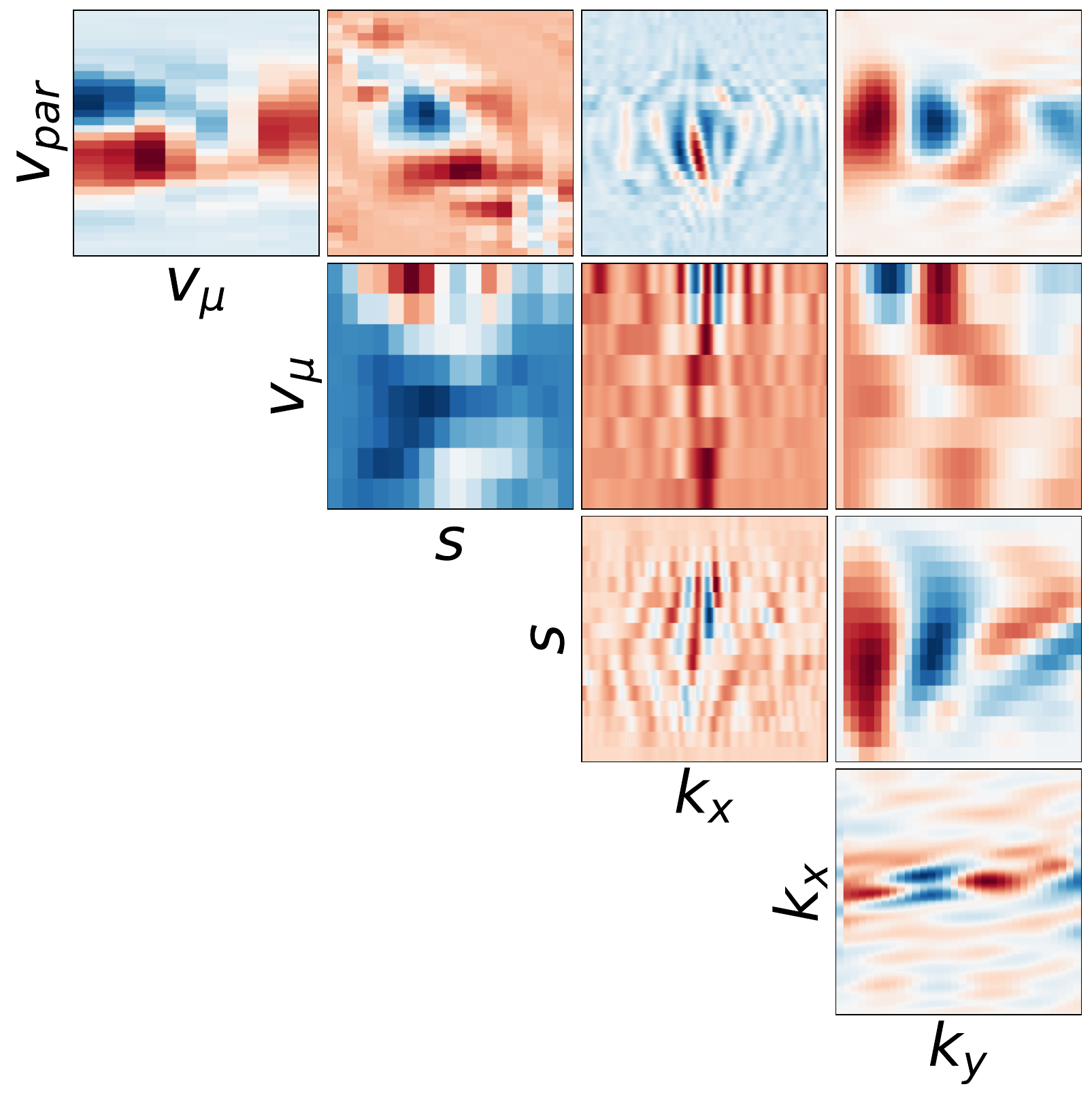}
        \caption{Middle slice}
        \label{fig:gt_middle_t0}
    \end{subfigure}
    \hspace{0.01\textwidth}
    \begin{subfigure}{0.3\textwidth}
        \includegraphics[width=\textwidth]{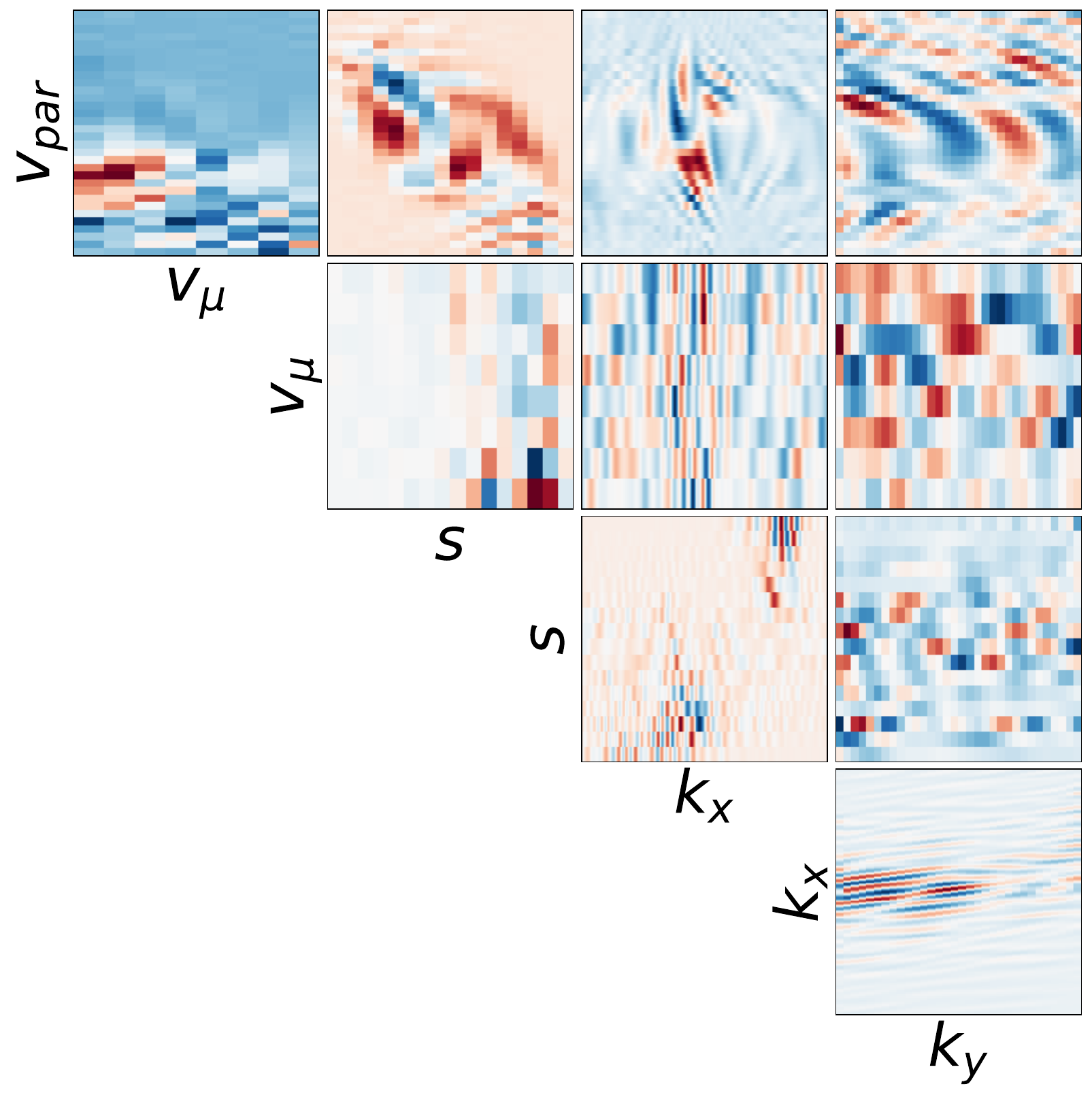}
        \caption{End slice}
        \label{fig:gt_end_t0}
    \end{subfigure}
    \caption{Distribution function $\delta f$ with an \ac{ITG} of $7.9$ at time $t=1.9$ (Linear Phase).}
    \label{fig:4x4_gt_t0}
\end{figure}

\begin{figure}[ht!]
    \centering
    \begin{subfigure}{0.45\textwidth}
        \includegraphics[width=\textwidth]{figs/5Dplots/compact_avg_7.9_50.6.pdf}
        \caption{Mean}
        \label{fig:gt_mean_t1}
    \end{subfigure}
    \hspace{0.05\textwidth}
    \begin{subfigure}{0.45\textwidth}
        \includegraphics[width=\textwidth]{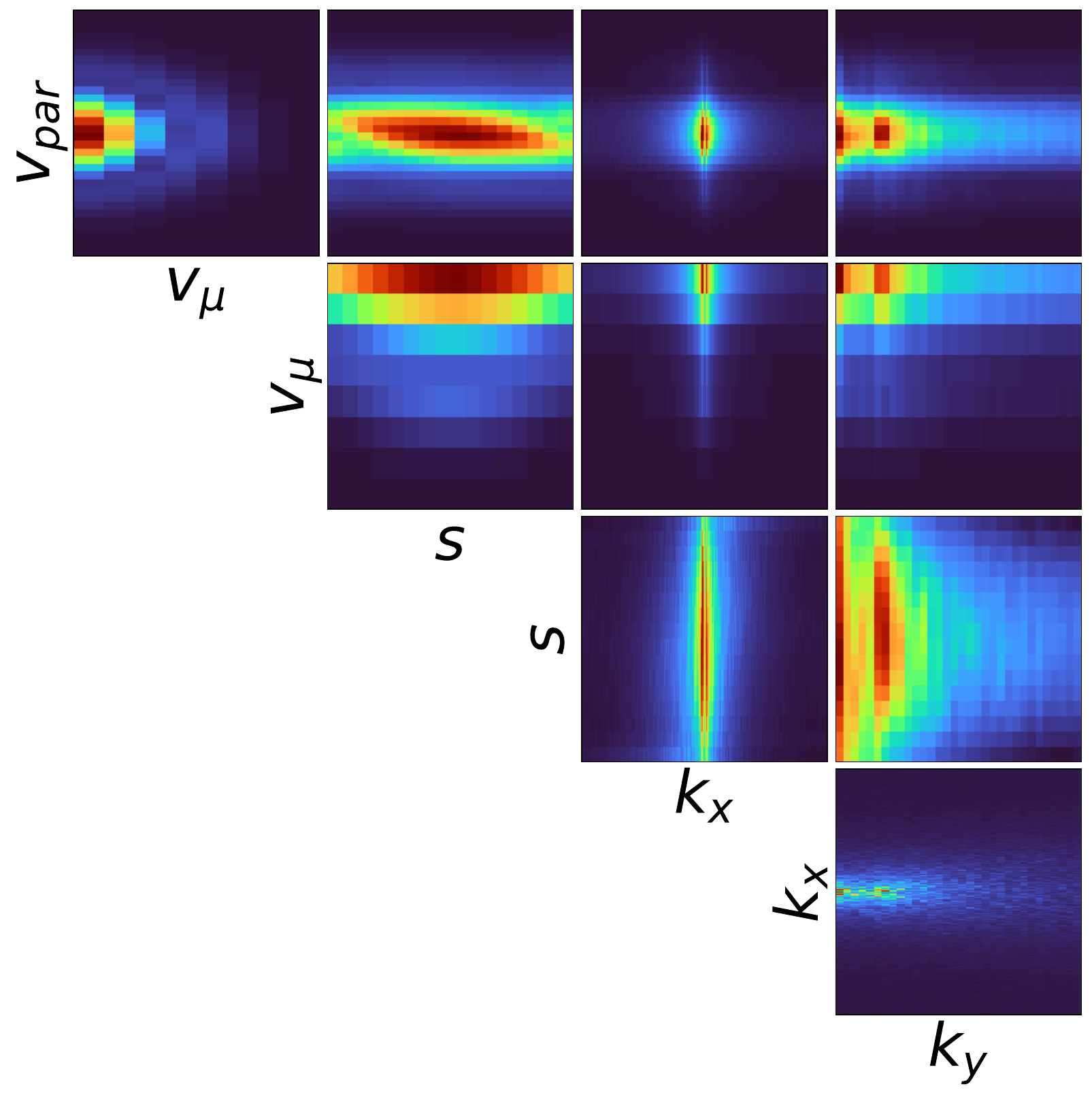}
        \caption{Standard Deviation}
        \label{fig:gt_std_t1}
    \end{subfigure}
    \par\bigskip %
    \begin{subfigure}{0.3\textwidth}
        \includegraphics[width=\textwidth]{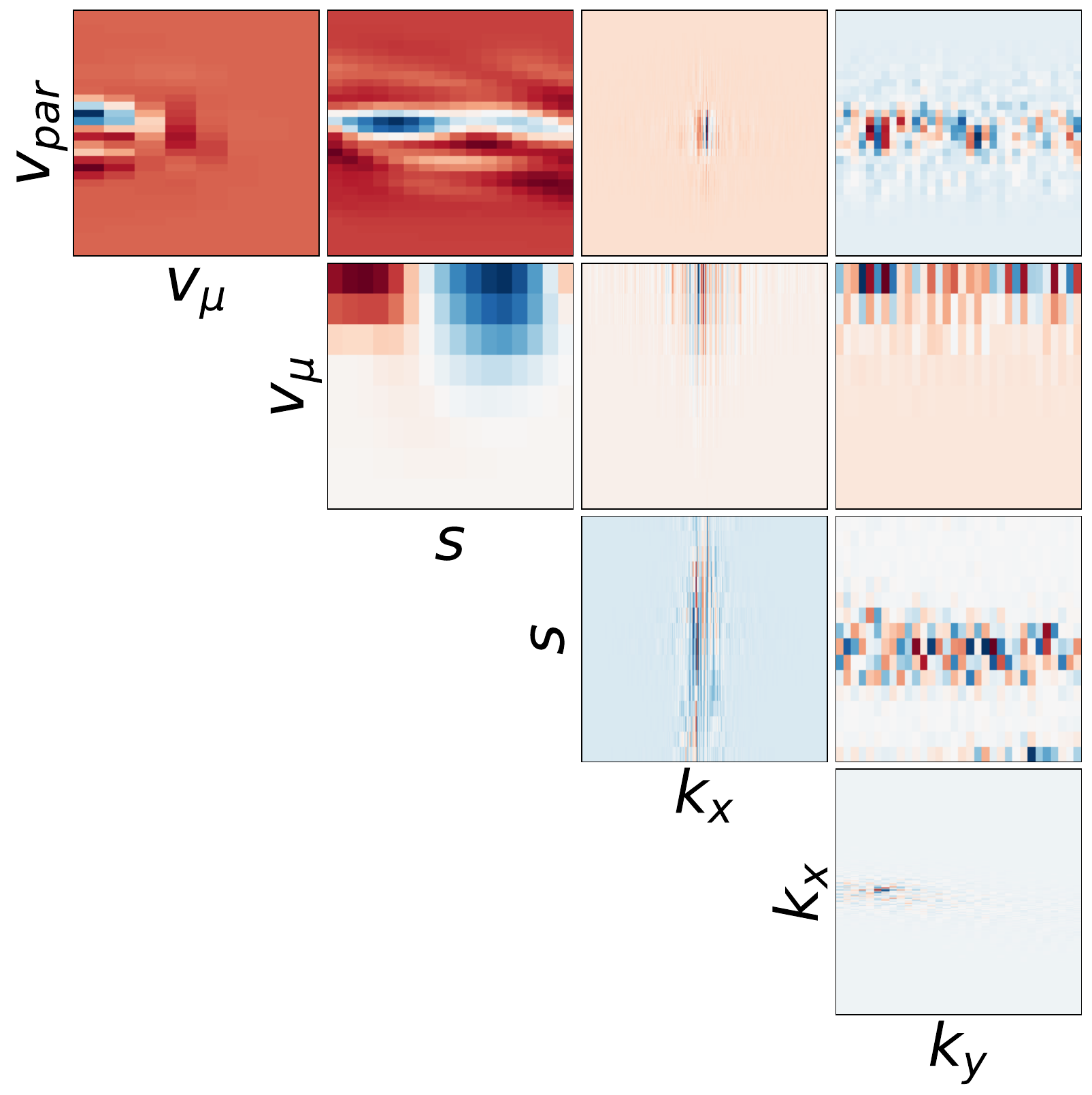}
        \caption{Start slice}
        \label{fig:gt_start_t1}
    \end{subfigure}
    \hspace{0.01\textwidth}
    \begin{subfigure}{0.3\textwidth}
        \includegraphics[width=\textwidth]{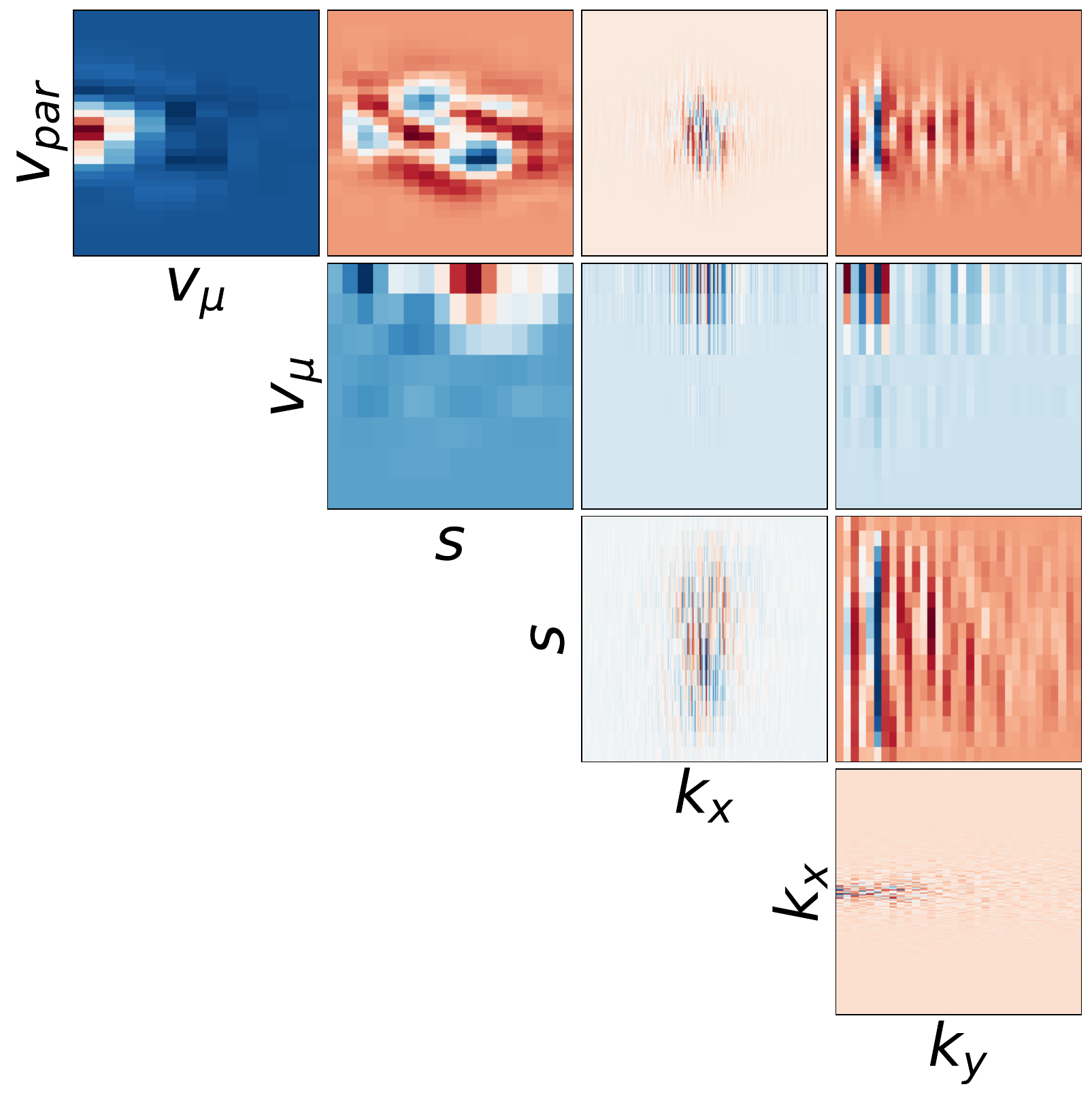}
        \caption{Middle slice}
        \label{fig:gt_middle_t1}
    \end{subfigure}
    \hspace{0.01\textwidth}
    \begin{subfigure}{0.3\textwidth}
        \includegraphics[width=\textwidth]{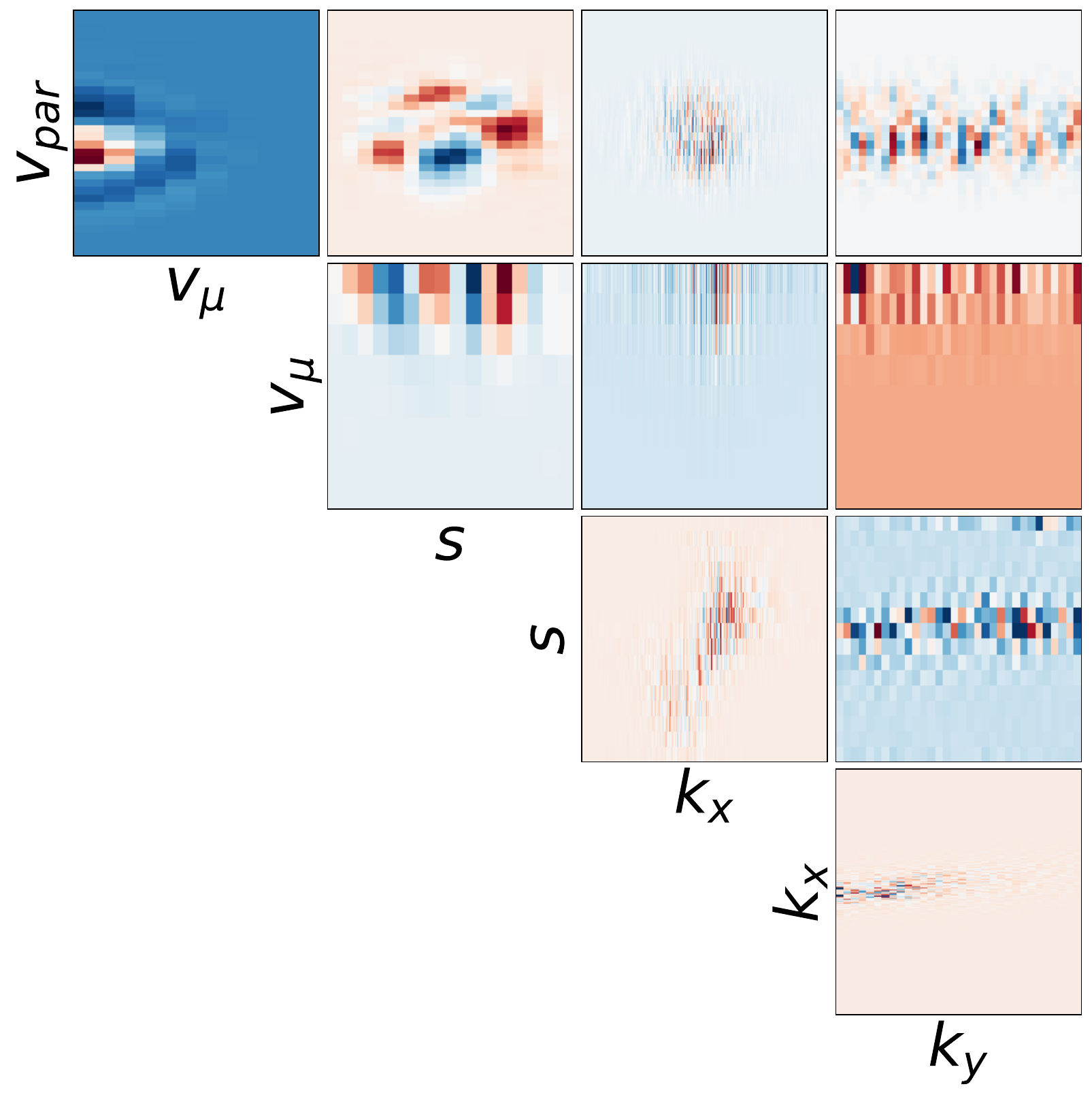}
        \caption{End slice}
        \label{fig:gt_end_t1}
    \end{subfigure}
    \caption{Distribution function $\delta f$ with an \ac{ITG} of $7.9$ at time $t=50.6$ (Saturated Phase).}
    \label{fig:4x4_gt_t1}
\end{figure}

\textbf{5D Data Visualization.}
We visualize an example of the generated data in \cref{fig:4x4_gt_t0} and \cref{fig:4x4_gt_t1}. 
They show the 5D distribution function for an \ac{ITG} of $7.9$ (green curve in \cref{fig:itg}) in the linear and saturated phases, respectively.
Those figures show all 10 pairwise combinations of axes.
To represent the remaining 3 axes in a 2D image, we compress them by taking the average (see \cref{fig:gt_mean_t0,fig:gt_mean_t1}), computing the standard deviation (see \cref{fig:gt_std_t0,fig:gt_std_t1}), or extracting slices at specific positions mainly at the start (see \cref{fig:gt_start_t0,fig:gt_start_t1}), in the middle (see \cref{fig:gt_middle_t0,fig:gt_middle_t1}), or at the end (see \cref{fig:gt_end_t0,fig:gt_end_t1}).

\section{Implementation details}
\label{app:impl_details}

While Vision Transformers usually employ convolutions for patch operations \citep{dosovitskiy_image_2021,liu_swin_2021,liu_video_2022}, we use simple fully connected layers because the computational cost of generalized convolutions can become significant in higher dimensions without an optimized implementation. 
Patch embedding, merging, and expansions are implemented as linear layers or MLPs.
The efficient application of \ndswinmsa{} is also implemented similarly to Swin. 
The shift is performed as a cyclic roll along all the window-partitioned axes. Because such operations can produce batches of windows that are not adjacent, we apply a n-dimensional window mask.

We use the Adam optimizer \citep{kingma_adam_2015} with a weight decay of 1e-5 and a cosine learning rate scheduler with linear warmups with a peak at 1e-3, decayed to 0.
During training we employ automatic mixed precision and gradient clipping to a magnitude of 1.
We also transfer the spectral domains ($k_x$ and $k_y$) to real space for training by applying inverse FFT.
Since a single training sample comprises several GBs of data, we perform lazy dataloading.
Model selection is performed every 20 training epochs based on the mean squared error on the holdout trajectory.

\end{document}